\documentclass[preprint2]{aastex}

\lefthead{Franceschini A. et al.}

\shorttitle{ISO investigates extremely-red hard X-ray sources}
\shortauthors{Franceschini et al.}

\begin{document}
\title{
 ISO investigates the nature of extremely-red hard X-ray sources responsible for
the X-ray background
\altaffilmark{1}}
\author{
A.\ Franceschini,$\!$\altaffilmark{2,3}
D.\ Fadda,$\!$\altaffilmark{4}
C. J.\ Cesarsky,$\!$\altaffilmark{5}
D.\ Elbaz,$\!$\altaffilmark{6}  
H.\ Flores,$\!$\altaffilmark{7}
G.L.\ Granato$\!$\altaffilmark{8}
}

\altaffiltext{1}{Based partly on observations obtained with the {\it Infrared
Space Observatory} and with {\it XMM-Newton}, ESA science missions with instruments 
and contributions funded by ESA Member States and the USA (NASA).}
\altaffiltext{2}{Dipartimento di Astronomia, Vicolo Osservatorio 2, I-35122 Padova, Italy}
\slugcomment{Astrophysical Journal Letters, submitted}
\altaffiltext{3}{Visiting Astronomer, European Southern Observatory, Garching, Germany}
\altaffiltext{4}{Instituto de Astrofisica de Canarias,
Via Lactea S/N, E38200 La Laguna (Tenerife), Spain}
\altaffiltext{5}{European Southern Observatory, Karl-Schwarzschild Strasse 2, 
D85748 Garching bei Munchen, Germany}
\altaffiltext{6}{CEA Saclay - Service d'Astrophysique, Orme des Merisiers, 
F91191 Gif-sur-Yvette Cedex, France}
\altaffiltext{7}{Observatoire de Paris Meudon, DAEC, F92195 Meudon Principal Cedex, France}
\altaffiltext{8}{Osservatorio Astronomico di Padova, Vicolo Osservatorio 2, 
I-35122 Padova, Italy}

\slugcomment{Astrophysical Journal Letters, submitted}

\begin{abstract}
We analyse very deep X-ray and mid-IR surveys in common 
areas of the Lockman Hole and the HDF North to study the sources of the X-ray background
(XRB) and to test the standard obscured accretion paradigm.
Observations with XMM-Newton and ISO of a substantial area in Lockman are particularly important 
to sample luminous - but relatively uncommon - obscured AGNs.
We detect a rich population of X-ray luminous sources with red optical colours, including 
a fraction identified with Extremely Red Objects (R-K$>5$) and galaxies with SEDs
typical of normal massive ellipticals or spirals at $z\sim 1$.
The X-ray luminosities of these objects ($L_{0.5-10 {\rm keV}}\sim 10^{43}-10^{45}$ erg/s) 
indicate that the ultimate energy source is gravitational accretion,
while the X-ray to IR flux ratios and the X-ray spectral hardness show evidence 
of photoelectric absorption at low X-ray energies.
An important hint on the physics comes from the mid-IR data at 6.7 and 15 $\mu$m,
well reproduced by model spectra of completely obscured quasars 
under standard assumptions and line-of-sight optical depths $\tau_{0.3\mu}\simeq 30-40$.
Other predictions of the standard XRB picture, like the distributions of intrinsic bolometric
luminosities and the relative fractions of type-I and -II objects (1:3), are also consistent 
with our results.
Obscured gravitational accretion is then confirmed as being responsible for the bulk of the X-ray 
background, since we detect in the IR the down-graded energy photoelectrically absorbed in X-rays:
63\% of the faint 5-10 keV XMM sources are detected in the mid-IR by Fadda et al. (2001). 
As discussed there, however, although as much as 90\% of the X-ray energy
production could be converted to IR photons, no more than 20\% (and possibly 
less) of the Cosmic IR Background can be attributed to X-ray loud AGNs.

\end{abstract}


\section{Introduction}
\label{intro}

The Cosmic X-ray (XRB) and Infrared (CIRB) Backgrounds are now well established
cosmological components, including a substantial fraction of the integrated
emissions by galaxies and AGNs over the Hubble time. 
While it is common wisdom that the XRB is largely due to past activity 
of AGNs with spectral properties determined by a wide range of column densities
for the line-of-sight absorbing gas (e.g. Setti \& Woltjer 1989; 
Madau et al. 1994; Comastri et al. 1995; Giacconi et al. 2001), the sources of 
the CIRB are less understood at present (Genzel \& Cesarsky 2000).

Although so far apart in photon energies, an important relationship has been
suggested to hold between these two components. The X-ray emission which is 
photoelectrically absorbed in type-II AGNs dominating the XRB  
is expected to be down-graded in energy by the dusty circum-nuclear medium 
and to emerge thermally re-processed in the
IR between a few and few hundreds $\mu$m (Efstathiou and Rowan-Robinson 1995; 
Granato, Danese, Franceschini 1997).
It has been even considered that, under rather extreme assumptions about the IR 
emissivity of type-II objects and the fraction of Compton-thick sources, 
half or so of the CIRB itself might be due to dust-reprocessed quasar emission 
(Almaini, Lawrence, and Boyle 1999; Fabian and Iwasawa 1999). 
If true, then both the XRB and partly also the CIRB would be manifestations 
of the same phenomenon of gas accretion in strong gravitational fields.

This simple, time-honored scheme for the origin of the hard XRB 
turned out difficult to prove, however. While
classical type-I quasars are easily detected in optical or soft X-rays at any
$z$, the putative type-II source population has remained elusive for long time,
confined by photoelectric absorption and dust extinction
into poorly sampled spectral domains -- the hard X-rays and the mid- and far-IR.
Tests of the type-II  population have been attempted, in particular, by 
correlating deep X-ray maps and catalogues of faint millimetric sources 
(e.g. Fabian et al. 2000; Hornschemeier et al. 2001; Barger et al. 2001). 
In general, however, X-ray sources are not detected in the millimeter, 
probably because of the lack of cold dust in these objects (the AGN heats the
circum-nuclear medium to temperatures too high to be observable at such long 
wavelengths even in high redshift sources).

New powerful instrumentation from space is eventually providing
opportunities for direct tests of the standard synthesis model of the XRB
and its obscured accretion paradigm.
More than 80\% of the XRB in the 2-8 keV band has
been resolved into sources with very long exposures by Chandra in the HDFN 
(Brandt et al. 2001) and the Chandra Deep Field South (Giacconi et al. 2001). 
XMM has resolved $\sim 60\%$ of the XRB in the harder band between
5 and 10 keV (Hasinger et al. 2001). For the first time representative samples 
of the sources of XRB are available for detailed physical inspection.

On the IR side, ISO has produced deep diffraction-limited
imaging between 12 and 18 $\mu$m (ISOCAM-LW3), allowing the detection
of hot dust emission by sources at large redshifts. 
%
ISO and {\sl Chandra} observations in a small field of the lensing
cluster Abell 2390 have been discussed by Wilman, Fabian and Gandhi (2000), with 
two sources found in common. Two other {\sl Chandra} sources in the HDF North with 
faint LW3 counterparts are reported in Alexander et al. (2001).
These X-ray/IR objects are quite faint and red in the optical, and lack 
spectroscopic redshifts (for one object Cowie et al. 2001 estimate z=1.47 from 
optical spectroscopy). Wilman et al. (2000) point out that the bright mid-IR flux
from two of these sources might be consistent with that expected from a 
circum-nuclear dust torus.
Though exploiting the deepest IR and X-ray data available, these results
are limited by the very small area of few tens square arcminutes in total.

We discuss in this paper the properties of a substantial population of
extremely red galaxies as counterparts of faint hard X-ray and IR sources,
based on combined deep observations by ISO and XMM of a 220 square
arcminutes field in the Lockman Hole by Fadda et al. (2001).
We exploit, in particular, their remarkable finding that a large fraction
(63\%) of the faint 5-10 keV XMM sources have relatively bright counterparts 
at 15 $\mu$m, already a hint that the IR detects in these sources the reprocessed 
energy absorbed in X-rays.  We also consider deeper data in a smaller area in the 
HDF North observed with {\sl Chandra}.
Sect. 2 summarizes the observational data on the samples.
In Sect. 3 we discuss the main physical properties of the sources based on
the available photometry and optical, IR and X-ray colours.
Sect. 4 is dedicated to a discussion and the conclusions.
We adopt H$_0$=70 Km/s/Mpc and $\Omega_m$=0.3, $\Omega_\lambda$=0.7.

\section{The sample}
\label{sample}

\subsection{Infrared, optical and X-ray data}

Our reference sample consists of 24 sources selected by XMM-Newton in the total
band 0.5-10 keV and detected by ISOCAM in a common region of 220 square arcmins located
in the Lockman Hole.    This area has been observed in the XMM-Epic PV phase 
(Hasinger et al. 2001) to flux limits of
$S_{0.5-2\ {\rm keV}}=3\ 10^{-16}$ erg/cm$^2$/s, 
$S_{2-10\ {\rm keV}}=1.4\ 10^{-15}$ erg/cm$^2$/s, and
$S_{5-10\ {\rm keV}}=2.4\ 10^{-15}$ erg/cm$^2$/s.
The high-energy XMM response allowed measuring X-ray hardness
ratios with small uncertainties: we will use in the following the
quantity HR=(H-S)/(H+S), where H=$S_{2-4.5\ {\rm keV}}$, S=$S_{0.5-2\ {\rm keV}}$.

\begin{figure}
\centerline{\plotone{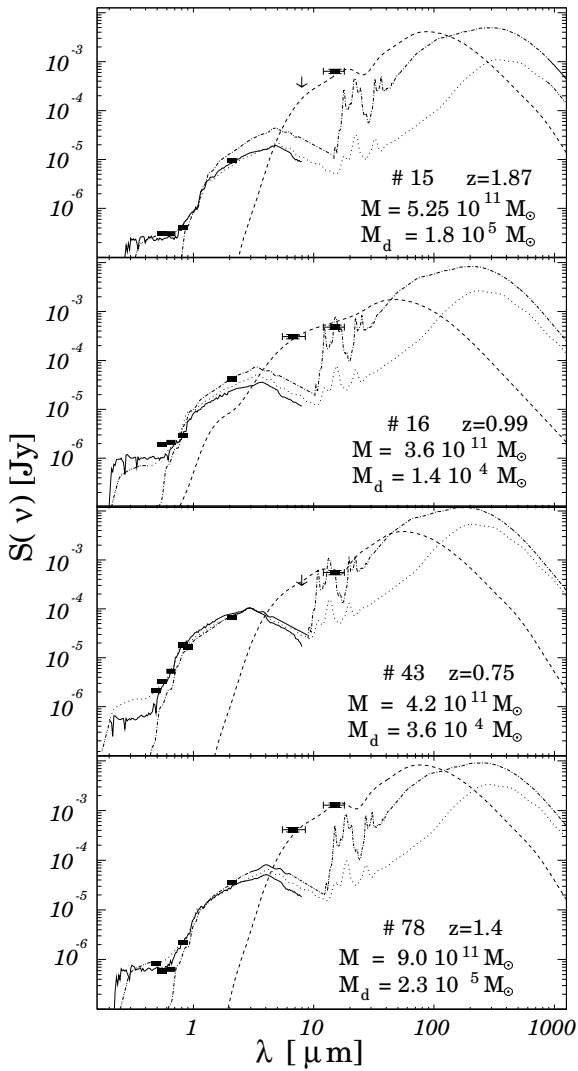} }
\vspace{0pt}
\caption{Spectral Energy Distributions of type-II X-ray AGNs found in the Lockman XMM/ISO
combined field (the$\#$ refers to the list in Fadda et al. 2001). 
Redshifts are photometric, but for source $\# 16$. 
Continuous line: our best-fit to the optical/NIR data
used for the photometric redshift. Dot-dashed line: SED of M82, used for comparison.
Dotted line: template of M51 fit to the NIR data to estimate the stellar mass M
in the host galaxy. 
Dashed line: model SED of dust emission by an obscured quasar fitting
the ISO 6.7 and 15 $\mu$m data (the circum-nuclear dust mass M$_d$ is also indicated).
Note that only for object $\# 16$ the IR emission from the quasar contributes to
the K-band flux.
}
\label{fig1}
\addtolength{\baselineskip}{1pt}
\end{figure}

The ISO observations of the Lockman Hole have been performed 
in the ISOCAM GT Program with the LW3 filter (12-18 $\mu$m) 
for a total of 60 ksec and with ISOCAM-LW2 (5-8.5 $\mu$m) for 70 ksec.
The 4$\sigma$ sensitivity limits are $S_{15\mu}\simeq 0.35$ mJy for the 
LW3 observations and $S_{7\mu}\simeq 0.3$ mJy for LW2.
The analysis of these data is reported by Fadda et al. (2001, 2002), who exploit 
them to evaluate the AGN contribution to the CIRB background. Twenty-two XMM sources 
are detected in LW3, two more in LW2 only, seven are detected in both LW2 and LW3. 
V, R, I and K magnitudes are reported, together with X-ray and IR data, 
in Fadda et al. (2001, their Table 3).

We also exploit a sample of 24 very faint {\sl Chandra} and ISOCAM sources 
in a small area of 25 square arcmin centered in the HDFN and detected by 
Brandt et al. (2001) and Aussel et al. (1999), with sensitivity limits of 
$S_{2-10\ {\rm keV}}=1.4\ 10^{-15}$ erg/cm$^2$/s and $S_{15\mu}\sim 0.05$ mJy.  
The cross-correlation of the two {\sl Chandra} and ISO catalogues is discussed in
Fadda et al. (2001, their Table 4).

\subsection{Redshift measurements}
\label{redshift}

Spectroscopic redshifts are available for all 24 HDFN sources but two, while
only 11 of the Lockman objects have spectroscopic measurements (mostly type-I quasars;
we operatively define a type-I object as having either blue optical colors or
broad-line emission in optical spectra; type-II objects are the complementary
population). For the remaining objects we estimated photometric redshifts with a tool
based on the PEGASE spectral synthesis code (Fioc and Rocca-Volmerange
1997). This allowed a large database of galaxy spectra to be synthesized, including 
effects of dust extinction. The tool has been trained on samples of ISO galaxies
in the HDF South with excellent results (Franceschini et al. in preparation).
For the typical SEDs of our sample objects, these estimates turned out to be rather
insensitive to the amount of extinction.
A comparison of our photometric estimates with Keck spectra
obtained for three Lockman galaxies (obtained as in Lehmann et al. 2001, and 
private communication) revealed the reliability of the procedure 
(errors $\Delta z\sim 0.1$, consistent with what found in HDF South).

Photometric redshifts for 11 Lockman and 1 HDFN sources have been obtained in this
way. Data on four type-II galaxies and best-fit solutions (thin continuous lines) 
are reported in Fig. 1.

\section{Properties of the X-ray/IR combined sample}
\label{properties}

\begin{figure}
\centerline{\plotone{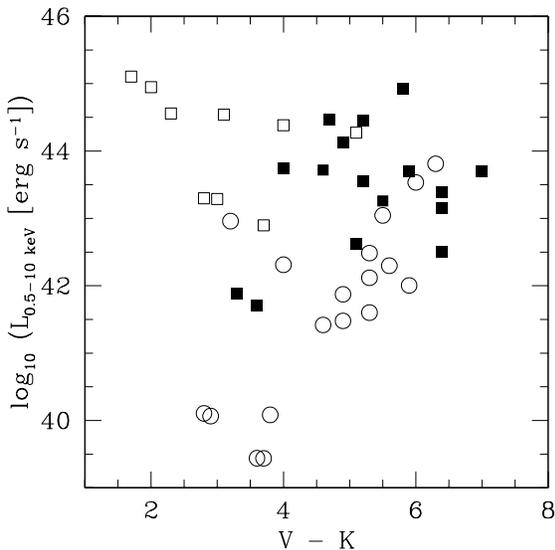} }
\vspace{0pt}
\caption{X-ray luminosity in the broad 0.5-10 keV band vs. V-K colours. 
Open squares: optically classified type-I AGNs, 7 coming from the Lockman and 2
from the HDFN samples. Filled squares: type-II AGNs from the Lockman (13 sources) 
and HDFN (3 sources). 
Type-II AGN classification in Lockman is from the present paper.
Open circles: unclassified {\sl Chandra} sources in the HDFN.
}
\label{fig3}
\addtolength{\baselineskip}{1pt}
\end{figure}

Figure \ref{fig3} is a plot of the X-ray luminosities versus V-K colours in the 
broad 0.5-10 keV band for the combined HDFN and Lockman samples. Evidently, the 
two cover different domains of the parameter space: the much fainter
HDFN data allow detection of low-luminosity emissions in moderate redshifts galaxies, 
part of which (those with $L_{0.5-10 {\rm keV}}\leq 10^{41}$ erg/s) should be attributed to 
stellar processes. Only in Lockman the survey area is large enough 
to include a sizeable number of high-luminosity sources.

\begin{figure}
\centerline{\plotone{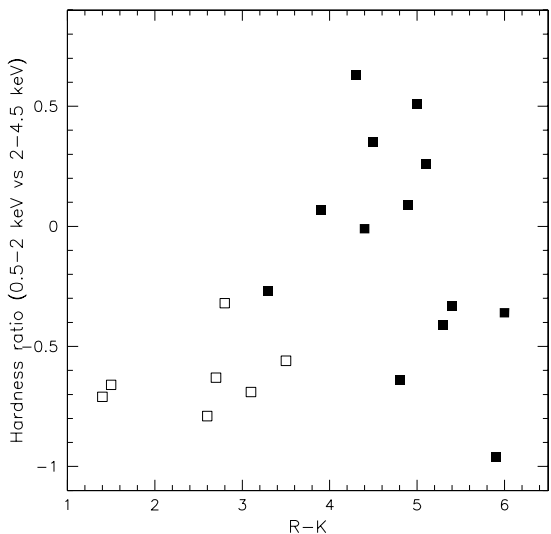} }
\vspace{0pt}
\caption{X-ray hardness ratios vs. R-K colours for the Lockman sources for which
the HR measure was possible. 
Same symbols as in Fig. \ref{fig3}.
}
\label{fig4}
\addtolength{\baselineskip}{1pt}
\end{figure}

The V-K colours for the combined sample are very widely spread between V-K$\simeq 1$ and
7. Objects classified as type-I QSOs based on existing optical spectra (open squares)
have blue colours (V-K$\leq 4$) and high luminosities. 
Instead, the bulk of the newly discovered XMM
sources (previously undetected in the ultra-deep ROSAT image of the field) have
very red optical SEDs (V-K$\geq 4$, filled squares in Fig. \ref{fig3}).
{\sl Several of these would be classified as Extremely Red Objects (EROs, R-K$\geq 5$ in Fig.
\ref{fig4}) by optical selection schemes} (e.g. Daddi, Cimatti, Renzini 2000).

Fig. \ref{fig4} shows a clear relationship between optical R-K colours and the
X-ray hardness ratios (HR): while the blue type-I objects have HR values
consistent with standard X-ray photon indices $\Gamma\sim 2$ ($HR\simeq -0.7$), 
{\sl the red sources display a wide range of HR values, including very hard X-ray spectra}.
Other {\sl Chandra} deep surveys have found similarly red, optically normal, galaxies 
hosting hard X-ray sources (Mushotzky et al. 2000; Fiore et al. 2000; Hornschemeier et 
al. 2001; Barger et al. 2001; see also Lehmann et al. 2001).

The important point illustrated by Fig. \ref{fig3} is that all the red ISO/XMM sources 
detected in Lockman have X-ray luminosities (L$_{0.5-10{\rm keV}}>10^{42}$ erg/s)
high enough to be classified as AGNs or quasars, rather than starbursts. For comparison,
none of the classical local starbursts (from M82 to Arp 220) approach these
luminosity regimes, all those exceeding it host an AGN (Fadda et al. 2001).

Assuming that an AGN dominates the X-ray production, Mushotsky et al. (2000) and Fiore 
et al. (2000) propose the following three different interpretations for the red population. 
{\sl (a)} AGNs hosting optically silent advection- or convection-dominated 
accretion flows (Ball et al. 2001, Di Matteo et al. 2000). 
{\sl (b)} A kind of red BL Lac objects. 
{\sl (c)} Heavily obscured AGNs. The last hypothesis, in particular, constitutes the basic 
assumption of standard models of the XRB (see Sect.1).
Our combined X/optical/mid-IR survey in Lockman offers a way to test these various 
hypotheses. 

In particular, we detect with ISO mid-IR emissions from a substantial fraction 
(33\%) of all Lockman XMM sources (Fadda et al. 2001). This fraction increases
to 63\% of the X-ray sources selected at very hard energies (5-10 keV).  
X-ray objects undetected in the 
IR are consistent with also belonging to the same population but fainter than the 
ISOCAM limits.
For the few sources for which we have both 6.7 and 15 $\mu$m flux detections (2 are 
reported in Fig. 
\ref{fig1}), the LW3/LW2 flux ratios are inconsistent with being due to a starburst, 
and require an AGN-like power source.

\begin{figure}
\centerline{\plotone{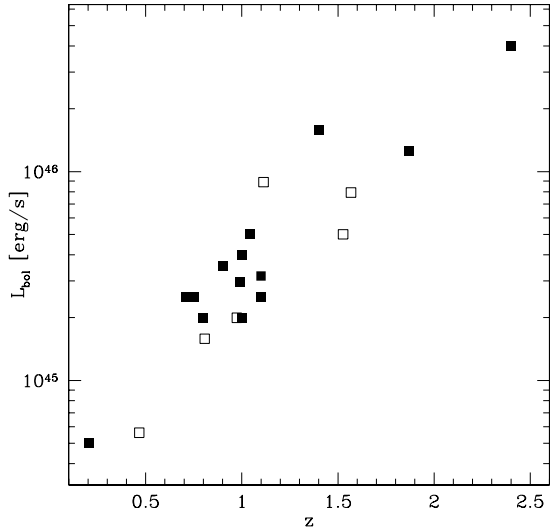} }
\vspace{0pt}
\caption{Bolometric luminosities of the underlying primary continuum for type-I 
(open squares) and type-II (filled squares) AGNs in Lockman detected by XMM and ISO.
}
\label{fig6}
\addtolength{\baselineskip}{1pt}
\end{figure}

We have compared in Fig. 1 predicted SEDs for dusty quasars (dashed lines) with our
observed optical/NIR/mid-IR data. We have used radiative transfer models
taken from Granato et al. (1997) and Andreani, Franceschini, Granato (1999), 
assuming a central point-like source surrounded by a dust distribution. The radius of the 
innermost dust shell is defined by the grain sublimation condition ($T\sim 1000$), 
which sets the
short-wavelength limit of dust emission, while the outermost radius is a free parameter affecting 
the spectrum at the longest wavelengths. Following Andreani et al. (1999), we have represented
a toroidal dust distribution as a "flared" disk with fixed covering factor $f=0.7$ and 
variable equatorial optical depth $\tau_{0.3\mu}$. Our observed SEDs for type-I QSO spectra 
are fit by solutions assuming pole-on ($\Theta=0^\circ$) or intermediate viewing inclinations.

Dashed lines in Fig. \ref{fig1} are solutions with $\Theta=90^\circ$
(edge-on view) and equatorial optical-depths of typically $\tau_{0.3\mu}\simeq 30-40$,
providing good fits to the observed 6.7 and 15 $\mu$m fluxes for type-II QSOs. 
An important result of these fits is that, given the very red optical spectra of these 
sources, the contribution by scattered 
light from the primary continuum should be minimal or absent (the parameter $\Theta$ in our 
model has to be $90^\circ$, see also Wilman et al. 2000).

We lack, unfortunately, sensitive sub-mm observations in the Lockman area to constrain the
long-wavelength tail of the spectra for the X-ray/IR AGN population: consequently, the 
maximum radius of the dust distribution and the dust masses are presently essentially
unmeasurable. On the contrary, data at 6.7 and 15 $\mu$m (in addition to those in optical/NIR
for type-I objects) provide a robust constraint on the bolometric source luminosity
(the assumption of a central illuminating source implies in particular that the the bulk 
of the energy is emitted between 5 and 40 $\mu$m for type-II QSOs). 
We report in Fig. \ref{fig6} as a function of $z$ our estimated bolometric luminosities 
between 0.1 and 1000 $\mu$m (the X-ray flux would only add a minor contribution, 
see Fig. 5 below) for type-I and -II AGNs in Lockman 
(open and filled squares respectively).
This plot indicates that there is no systematic difference between the bolometric emissions 
of the two classes, at the various redshifts. This is in agreement with the standard
unification paradigm.

\begin{figure}
\centerline{\plotone{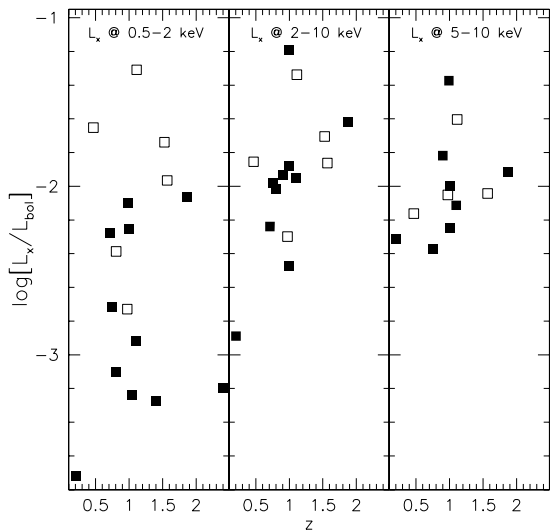} }
\vspace{0pt}
\caption{Fractions of X-ray luminosity over the bolometric luminosity for type-I (open squares)
and type-II (filled squares) AGNs in Lockman. Various panels refer to luminosities in different
X-ray bands (as indicated).
}
\label{fig5}
\addtolength{\baselineskip}{1pt}
\end{figure}

The X-ray flux has, on the contrary, a sensitive dependence on the column density of the 
obscuring medium.   Fig. \ref{fig5} reports the fraction of radiation emitted in the various 
X-ray bands over the bolometric emission, and shows that type-II QSOs are systematically
fainter emitters than type-I in soft X-rays, the difference reducing in the harder
bands. Figs. \ref{fig4} and \ref{fig5} require column densities of the X-ray absorbing gas
for type-II objects in the range $N_H\simeq 10^{22}$ to $10^{23}$ cm$^{-2}$, not inconsistent
with the extinction of $\tau_{0.3\mu}\simeq 30-40$ inferred from the fits to the IR spectra.
At the same time, Fig. \ref{fig5} implies that the contribution of the X-ray energy
to the bolometric luminosity cannot be dominant: the bulk of the QSO primary energy
is produced in the UV-optical, consistent with the average type-I spectrum by Elvis et al.
(1994).

Finally, the completeness and large mid-IR identification fraction of our XMM 
sample in Lockman allow also a statistical test of XRB models.
Again in rough agreement with the unification scheme (e.g. Lawrence 1991),
the ratio of the number of type-II to type-I QSOs appearing in Fig. \ref{fig6} is 
$\simeq 3$ in the redshift interval $0.5<z<1.5$ of maximal sensitivity for our 
combined X/IR survey.

\section{Conclusions}
\label{conclusions}

We have exploited very deep X-ray and mid-IR survey data obtained for the first time
by ISO, XMM-Newton and {\sl Chandra} in common areas of the Lockman Hole
and HDFN, to test standard models for the origin of the XRB and the supposed dust-obscured
accretion phenomenon.
The large surveyed area in Lockman was essential to detect significant numbers of luminous 
obscured AGNs, too rare to be appropriately sampled in the small HDFN field.

This combined X-ray/IR survey detects normal type-I quasars with standard optical, IR and
X-ray properties (blue colours, power-law X-ray and IR spectra). 
In addition, a rich population of X-ray luminous sources with red
optical colours is identified, roughly half of which would be classified as Extremely
Red Objects (R-K$>5$) in the optical. Their optical SEDs are those typical of normal massive
elliptical or spiral galaxies at the appropriate $z$.
X-ray sources with similarly red counterparts have also been occasionally reported 
(Fiore et al. 2000; Mushotzky et al. 2000; Barger et al. 2001; 
Lehmann et al. 2001; Cowie et al. 2001).

While the ultimate nuclear energy source in these objects has to be quasar-like 
gravitational accretion,
given the large X-ray luminosities ($L_{0.5-10 {\rm keV}}\sim 10^{43}-10^{44}$ erg/s),
an important hint on the physics comes from the mid-IR data at 6.7 and 15 $\mu$m.
These data are well reproduced by model spectra of obscured quasars 
under standard assumptions and line-of-sight optical depths of typically 
$\tau_{0.3\mu}=30-40$.
A detailed spectral analysis, including radiative transfer in the IR and comptonization 
effects in the X-rays, is in progress (Granato et al. in preparation).

Altogether
we find that various predictions of the standard XRB picture (e.g. Madau et al. 1994;
Comastri et al. 1995) are met by our analysis, within the uncertainties implied by 
the small number of sources.  In particular, 
the bolometric luminosity distributions do not appear to be systematically different
between type-I and -II objects, and their observed relative fractions are consistently 
close to the canonically assumed value of 1:3. 
Also the X-ray luminosities and hardness ratios of type-II objects show evidence 
of photoelectric absorption at low X-ray energies.

We believe that these results provide important new evidence that obscured gravitational 
accretion in massive normal-looking galaxies is responsible for the bulk of the X-ray 
background: the energy which is absorbed in X-rays (which could be as much as 80-90\% when
averaged over the whole population producing the XRB, see Fabian \& Iwasawa 1999) is found 
by us to be down-graded in photon energy and re-emitted in the IR, as expected. 
In spite of this, and following the correlation analysis of deep X-ray and IR images
by Fadda et al. (2001), 
probably no more than 20\% (and possibly much less, see Elbaz et al. 2001) 
of the Cosmic IR background can be attributed to X-ray loud AGNs.

\acknowledgements
A.F., D.F. and H.F. thank the European Southern Observatory 
for hospitality during elaboration of this work. 
Research supported by EU RTN Network "POE" HPRN-CT2000-00138.

\end{document}